\documentclass[prd,showpacs,nofootinbib,preprintnumbers,floatfix]{revtex4}  
\usepackage[plainpages=false, colorlinks=true, anchorcolor=blue, linkcolor=blue, citecolor=blue, bookmarks=false]{hyperref}
\usepackage{graphicx}  
\usepackage{dcolumn}   
\usepackage{bm}       
\usepackage{amssymb}   
\usepackage{natbib}
\usepackage{amsmath}  
\usepackage{longtable}  
\usepackage{float} 
\usepackage[utf8]{inputenc}
\usepackage{tasks}
\usepackage{booktabs} 
\usepackage{array} 
\usepackage{tikz}
\usetikzlibrary{shapes,arrows,positioning,automata,backgrounds,calc,er,patterns,scopes}
\usetikzlibrary{arrows.meta}
\usetikzlibrary{decorations.markings}
\DeclareMathAlphabet{\pazocal}{OMS}{zplm}{m}{n}

\begin{document}

\newcommand{\apjl}{Astrophys. J. Lett.}
\newcommand{\apjs}{Astrophys. J. Suppl. Ser.}
\newcommand{\aap}{Astron. \& Astrophys.}
\newcommand{\rthis}[1]{\textcolor{black}{#1}}
\newcommand{\aj}{Astron. J.}
\newcommand{\pasp}{PASP}
\newcommand{\araa}{Ann. Rev. Astron. Astrophys. } 
\newcommand{\aapr}{Astronomy and Astrophysics Review}
\newcommand{\ssr}{Space Science Reviews}
\newcommand{\mnras}{Mon. Not. R. Astron. Soc.}
\newcommand{\apss} {Astrophys. and Space Science}
\newcommand{\jcap}{JCAP}
\newcommand{\na}{New Astronomy}
\newcommand{\pasj}{PASJ}
\newcommand{\pasa}{Pub. Astro. Soc. Aust.}
\newcommand{\physrep}{Physics Reports}

\title{\boldmath Cosmological bounds on a possible electron-to-proton mass ratio variation and constraints in the  lepton specific 2HDM}
\author{R. G. Albuquerque$^1$}\email{rennan.gomes@ufabc.edu.br}

\author{R. F. L. Holanda$^{2,3}$}\email{holandarfl@fisica.ufrn.br}

\author{I. E. T. R. Mendon\c{c}a$^{2,5}$}\email{isaacmendonca@biofy.tech}

\author{P. S. Rodrigues da Silva$^{1}$}\email{psilva@fisica.ufpb.br}

\affiliation{$^1$Departamento de F\'{\i}sica, Universidade Federal da Para\'\i ba, Caixa Postal 5008, 58051-970,
Jo\~ao Pessoa, PB, Brasil}
\affiliation{$^2$Departamento de F\'{\i}sica, Universidade Federal do Rio Grande do Norte,Natal - Rio Grande do Norte, 59072-970, Brasil}
\affiliation{$^3$Departamento de F\'{\i}sica, Universidade Federal de Campina Grande, 58429-900, Campina Grande - PB, Brasil}
\affiliation{$^5$Biofy.tech IA Department, 38407-180, Uberlândia-MG Brazil}

\begin{abstract}
In this work, we test a possible redshift variation of
the electron-to-proton mass ratio, $\mu = m_e/m_p$, directly from galaxy cluster gas mass fraction measurements and type Ia supernovae observations. Our result reveals no variation of $\mu$ within 1~$\sigma$. From the point of view of Particle Physics, we can use the precision on these results to constrain the parameter space of models beyond the Standard Model of electroweak interactions. 
We exemplify this by focusing on a specific Two-Higgs doublet model (2HDM), where the second scalar doublet couples exclusively to leptons. 
An important parameter in the model concerns the ratio between its vacuum expectation values, defined by $\tan\beta\equiv v_2/v_1$. In our approach, we can constrain the inverse parameter $(\cot\beta)$ to an optimal value, $(\cot\beta)= (2.003 \pm 0.081)\cdot 10^{-3}$, with the highest vacuum expectation value for 2HDM, $v_2$, estimated at around $240.57 \pm 2.93$~GeV. Also, by taking into account the discrepancy in the anomalous magnetic moment of the muon found between theory and experiment, we can reduce the validity region for this model and establish bounds on the scalar masses, in light of our findings from galaxy cluster data for $\mu$. This study contributes valuable insights to the understanding of the interface between Particle Physics and Astrophysics, establishing a new interrelationship between data on the large-scale structure of the Universe and subatomic Physics.
\end{abstract}

\maketitle
\flushbottom

\section{Introduction}
\label{sec:intro}

Fundamental physical theories rely on certain parameters known as fundamental constants, such as the fine-structure constant, gauge couplings, and Yukawa couplings. Their values cannot be predicted from existing knowledge and must be determined through experimentation. These constants are traditionally assumed to be mathematical constructs that are unchanging across both time and space, as posited by the simplest and most successful physical theories. Dirac \cite{Dirac1937} was among the first to question whether these constants are merely mathematical constructs or if they hold deeper significance within a broader cosmological framework. Current laboratory experiments have ruled out significant variations in these constants on  laboratory \cite{Dzuba1999}, geological \cite{Fujii2000,Flambaum2006}, and solar system time scales \cite{Olive2004}. Astrophysical methods, particularly the spectroscopy of quasar absorption lines, which involves measuring transition wavelengths at high redshifts over vast time intervals, prove a much broader time span \cite{Fiorenzano2004,Tzanavaris_2005,Molaro2013,Murphy2016}.

In recent years, observational tests investigating possible variations in fundamental constants have proliferated in the literature (see \cite{Uzan2003,Barrow_2005,Martins_2017,Bambi2022} for a review). In this context, two dimensionless constants, the fine-structure constant ($\alpha \equiv e^{2} / 4\pi \epsilon_0 \hslash c$) and the electron-to-proton mass ratio ($\mu \equiv m_e/m_p$), are of particular importance. These constants are not only interrelated, but their potential variation is also connected to new physics, including models of extended gravity~\cite{Wei2011,Harko2013,Hess2014}, string theory~\cite{Damour1994,Martins2015}, and more recently dark matter models~\cite{Davoudiasl2019,Brzeminski2021}. Thus, constraining the temporal and spatial variations of these parameters could have profound implications for both cosmology and fundamental physics.
The use of large-scale structure physics, particularly the physics of galaxy clusters, to constrain variations in $\alpha$ is not new~\cite{Galli_2013}; however, it has become an area of intense investigation in recent years~\cite{Martino2016a,Martino2016b,rodrigo_leonardo,2016JCAP...08..055H,2019JCAP...03..014C,Liu_2021}. In contrast,  no studies have yet addressed the issue of $\mu$ variation using galaxy clusters, a path that we aim to explore in this work.

Concerning small-scale physics, the Standard Model of Electroweak Interactions~\footnote{ We will neglect the strong interaction sector of the Standard Model of Particle Physics in this work, the $SU(3)_C$ part of the gauge symmetry group, as it is not important for the entire analysis we deal with.} (SM) has been thoroughly confirmed over the past decades through several collider data measurements, culminating in the discovery of the last important ingredient predicted by the model, the Higgs boson, at the Large Hadron Collider (LHC)\cite{ATLAS:2012yve,CMS:2012qbp,ATLAS:2022vkf}. This was a significant step in definitively establishing the SM as a fiducial paradigm for describing the visible matter content in the Universe and its interactions in a detailed and precise way. However, it is not the most complete theory we can imagine. This is because there are several unresolved issues, such as the evidence for Cold Dark Matter\cite{Bertone:2016nfn}, the mass and oscillation of neutrinos~\cite{deSalas:2020pgw,Capozzi:2017ipn}, and the experimental deviation from the theoretical prediction of the muon anomalous magnetic moment~\cite{muon(g-2),muon(g-2)/2}, among others, that cannot be addressed by the model without extending its matter or interaction content (gauge group). In this sense, several models have been proposed in the literature to go beyond the SM. A particular class of interest for phenomenologists is the so-called Two-Higgs Doublet Model (2HDM)\cite{2HDM_2012,Wang:2022yhm}, especially since the scalar sector of the SM is under intense scrutiny. An extra scalar doublet represents the simplest component of supersymmetric theories\footnote{Supersymmetric theories (for a review, see Ref.~\cite{Martin:1997ns,Luty:2005sn,Mummidi:2023ysa} and references therein) have been extensively investigated for decades, as they are a fundamental component for the mathematical consistency of string theory. However, no evidence of supersymmetry has been observed in accelerators to date.} that can be studied theoretically without necessarily addressing the specific details of the intricate spectrum of such theories. In addition, 2HDM can introduce sources of CP violation in the scalar sector~\cite{Lee:1973iz}, allowing baryogenesis in this class of models~\cite{Bochkarev:1990fx,Turok:1990zg}. The extra neutral scalar can also offer a richer vacuum structure that could lead to a first-order phase transition~\cite{Cline:1996mga,Fromme:2006cm,Cline_2011,Dorsch:2013wja}, another requirement for baryogenesis. Finally, it is well known that additional scalar doublets do not alter the SM prediction for the observable $\rho$-parameter\footnote{The $\rho$-parameter is defined (at tree level) as $\rho_{\text{tree}} = \frac{M_W^2}{M_Z^2 \cos^2 \theta_W}$, where $M_W$ and $M_Z$ are the masses of the W and Z weak gauge bosons, respectively, and $\theta_W$ is the weak mixing angle. Its theoretical value in the SM is $\rho_{\text{tree}} = 1$. }.

Our main goal in this work is to constrain a possible redshift variation in the electron-to-proton mass ratio, $\mu\equiv m_e/m_p$, using observational data, specifically galaxy cluster gas mass fraction measurements and Type Ia supernova luminosity distances, and to impose limits on particle physics models beyond the Standard Model (SM). We focus on a specific version of the 2HDM, where leptons and quarks interact with different scalar doublets. Since the constraint on $\mu$ from galaxy clusters is directly related to the ratio between the vacuum expectation values (vev) of the model~\footnote{As we will see, in our approach we restrict the variation in the electron-to-proton mass ratio to the electron mass only. This restriction allows us to infer that the ratio of vev in the 2HDM, namely $\tan{\beta}$, is the parameter constrained by the $\mu$ variation.}, we can use these constraints on $\mu$ to derive bounds on the parameters of the 2HDM. To impose tighter limits on the parameter space, including the scalar masses of the model, we also compute the anomalous magnetic moment of the muon, restricting the analysis to the region of parameters that can resolve the $4.2\sigma$ discrepancy between experiment and theory. This discrepancy currently represents one of the most promising indications of physics beyond the SM~\cite{Abi_2021}.

The manuscript is organized as follows: we first present the methodology by which we derive the relation between the electron-to-proton mass ratio and the galaxy cluster gas mass fraction in Section~\ref{methodology}. Then, in Section~\ref{data}, we briefly explain the cosmological data sample used in our analysis. We describe, in Section~\ref{2HDMsec}, the theoretical structure of the 2HDM that we base our investigation. In Section~\ref{sec:analysis} we describe the details of our analysis and the results we get for the $\mu$ variation. After that, in Section~\ref{sec:muon-anomaly}, we present constraints on $\tan{\beta}$, as well as the masses of the heavier Higgs in the 2HDM, including the computation of the muon anomalous magnetic moment contribution to solve the theoretical-experimental discrepancy. Finally, we draw our conclusions and share some perspectives in Section~\ref{sec:conclusions}.

\section{Electron-to-proton mass ratio and galaxy cluster  gas mass fraction}
\label{methodology}

Galaxy clusters are the largest virialized structures in the universe \cite{2011ARA&A..49..409A}. They are predominantly composed of dark matter and hot intergalactic gas, referred to as the Intra-Cluster Medium (ICM). The ICM, in a plasma state, emits a continuous X-ray spectrum primarily dominated by thermal bremsstrahlung, with typical temperatures ranging between $2$ and $10 \, \mathrm{keV}$ \cite{Sarazin1988}.
 The ratio of ICM mass to total cluster mass, that is, the gas mass fraction \(f_{gas}\), is of particular significance in cosmology, as it allows constraints on cosmological parameters \cite{Allen2008,Ettori2009,Mantz2014}. Moreover, it enables various tests on the fundamental physics, including, for instance, the exploration of fundamental constant variations \cite{rodrigo_leonardo, eu_rodrigo2} and violations of Einstein's equivalence principle \cite{eu_rodrigo1,2017CQGra..34s5003H,2017PhRvD..95h4006H}. Consider the gas mass, \(M_{gas}\), within a radius \(R\),  obtained through X-ray spectrum measurements \cite{Sarazin1988,Sasaki1996}:
\begin{equation}
\begin{aligned}
M_{gas}(<R) &= \left( \frac{3 \pi \hbar m_e c^2}{2 (1+X) e^6} \right)^{1/2} 
\left( \frac{3 m_e c^2}{2 \pi k_B T_e} \right)^{1/4} M_H \\
&\times \frac{1}{[\overline{g_B}(T_e)]^{1/2}} {r_c}^{3/2} 
\left[ \frac{I_M (R/r_c, \beta)}{I_L^{1/2} (R/r_c, \beta)} \right] [L_X (<R)]^{1/2}.
\end{aligned}
\label{eqn_Mgas}
\end{equation}
Here, \(X\) is the fraction of Hydrogen mass in the gas, \(m_{e}\) and \(M_H\) are the masses of the electron and Hydrogen~\footnote{We designate the mass of Hydrogen with a capital M to avoid confusion with the mass of the heavy Higgs to be defined later in the article.}, respectively. \(T_{e}\) is the gas temperature, \(\overline{g}_{B}\) is the Gaunt factor, a multiplicative correction to take quantum effects into consideration for the calculations of absorption or emission of radiation~\cite{dopita2003astrophysics}, and \(r_{c}\) is the core radius. \(I_{M}\) and \(I_{L}\) are defined as:
$$
I_M (y, \beta) \equiv \int_0^y (1+x^2)^{-3 \beta/2} x^2 dx,
$$
$$
I_L (y, \beta) \equiv \int_0^y (1+x^2)^{-3 \beta} x^2 dx,
$$
obtained using a spherical $\beta$ model, and the bolometric luminosity \(L_{X}(<R)\) is given by:
$$
L_{X}(<R) = 4\pi [D_{L}(z, \Omega_{0}, H_{0} )]^{2}f_{X}(< \theta)\,\,\, ,
$$
where \(f_{X}(<\theta)\) is the total bolometric flux within the outer angular radius \(\theta\)~\cite{Peebles1983}. \(D_{L}\) is the luminosity distance, which depends on the redshift, $z$, the cosmological density parameter, $\Omega_0$,  and the Hubble parameter, $H_0$. The total cluster's mass, \(M_{tot}\), obtained by using the Euler equation for a fluid in hydrostatic equilibrium, with the additional assumption of isothermality~\cite{white2011fluid}, is given by:
\begin{equation}
M_{tot}(<R) = - \left. \frac{k_B T_e R}{G x_{H} M_H} \frac{d \ln n_e(r)}{d \ln r} \right|_{r=R}.
\label{eqn_Mtot}
\end{equation}
Here, \(x_{H}\) is the molar fraction of hydrogen in the gas, $G$ is Newton's gravitational constant, and \(n_{e}\) is the electron gas density. Thus, from Eq.~(\ref{eqn_Mgas}), we observe that \(M_{gas} \propto M_H\), and from Eq.~(\ref{eqn_Mtot}), \(M_{tot} \propto 1/M_H\). Consequently, we can conclude that the gas mass fraction, $f_{gas}$ within the intra-cluster medium, which is the ratio between Eq.~(\ref{eqn_Mgas}) and Eq.~(\ref{eqn_Mtot}),
\begin{equation}
f_{gas} \propto M_H^{2} ,
\end{equation}
and since \(M_H = m_{e} + m_{p}\)~\footnote{Here we neglect the binding energy of the hydrogen atom, since it is of the order of $\sim 10^{-4}$ of the electron mass.},  then
\begin{equation}
f_{gas} \propto \left(1 + \mu\right)^{2}m_{p}^{2},
\label{eqn_frac_gas_mod}
\end{equation}
which, as we can see, explicitly depends on \(\mu\), the ratio of electron to proton masses. 

On the other hand, the expected constancy of $f_{gas}$ within massive, hot and relaxed galaxy clusters with redshifts in the interval of interest has been used to constrain cosmological parameters via the following equation \cite{Allen2008,Ettori2009,Mantz2014}:
\begin{equation}
f_{gas}^{th}(z) = \gamma(z) \, K_0, A(z)  \frac{\Omega_{b}}{\Omega_{m}}\left[\frac{D^{ref}_{L}}{D_{L}}\right]^{3/2}.
\label{eqn_pheno_fgas}
\end{equation}
In Eq.~(\ref{eqn_pheno_fgas}), \(D_{L}^{ref}\) represents the luminosity distance of the cluster calculated using a fiducial concordance model (the flat \(\Lambda \text{CDM}\) model with \(H_0=70\) and \(\Omega_m=0.3\)), while \(D_{L}\) is the actual luminosity distance of the cluster. The term \(\gamma(z)\) is the baryon depletion factor, expressed as
\[
\gamma(z) = \gamma_0(1+\gamma_1 z)\left( \frac{M_{2500}}{3 \times 10^{14} M_{\odot}} \right)^\alpha,
\]
where \(\alpha = 0.025 \pm 0.033\), \(\gamma_0 = 0.79 \pm 0.11\) and \(\gamma_1 = 0.07\). The last two values were obtained from hydrodynamical simulations in Refs.~\cite{10.1093/mnras/stt265, Mantz_2021, 10.1093/mnras/stt2129}. The parameters \(\Omega_b\) and \(\Omega_m\) represent the baryonic and matter density parameters, respectively. The term \(K_0 = 0.96 \pm 0.12\) refers to the mass calibration factor, which was determined using 13 clusters from the sample, incorporating weak gravitational lensing measurements from the Weighting the Giants project \cite{2016MNRAS.457.1522A}. The \(A(z)\) term represents the angular correction factor, defined as 
\[
A(z) = \frac{H(z) \, d(z)}{\left[H(z) \, d(z)\right]^\mathrm{ref}},
\]
where \(H(z) = \sqrt{\Omega_m (1+z)^3 + (1-\Omega_m)}\), and \(d(z)\) is the comoving distance to the cluster (here $\Omega_m=0.315\pm 0.007$ and $H_0=67.36 \pm 0.54 $ km/s/Mpc since we have considered the flat $\Lambda$CDM model \cite{Planck2020}).

Thus, considering a possible variation of \(\mu\) with redshift, we modify Eq.~(\ref{eqn_pheno_fgas}) accordingly, arriving at the expression:
\begin{equation}
f_{gas}^{obs} = \left(1+\mu(z)\right)^{2} \gamma(z) \, K_0\, A(z) \frac{\Omega_{b}}{\Omega_{m}}\left[\frac{D^{*}_{L}}{D_{L}}\right]^{3/2}.
\end{equation}
Finally, if the luminosity distance is known for each galaxy cluster, \(\mu(z)\) can be inferred through the deviation from zero of \((1+\mu)^2\) with respect to the ratio
\[
\frac{f_{\text{gas}}^{\text{obs}}}{f_{\text{gas}}^{\text{th}}},
\]
where \(f_{\text{gas}}^{\text{obs}}\) is the observed gas mass fraction and \(f_{\text{gas}}^{\text{th}}\) is the theoretical gas mass fraction.
\begin{equation}
\label{fgas_mu}
\left(1 + \mu \right)^{2} - \frac{f_{\text{gas}}^{\text{obs}}}{f_{\text{gas}}^{\text{th}}} = 0 \,\,\, , 
\end{equation}
where $f_{\text{gas}}^{\text{obs}}$ is the gas mass fraction dataset\footnote{Which would be equivalent to the modified equation for $f_{gas}$.} presented in Section \ref{data} and $f_{\text{gas}}^{\text{th}}$ comes from Eq.~(\ref{eqn_pheno_fgas}). This will provide 40 values of $\mu$ (see Table II), which will be used to obtain the expected variation in the range of redshifts presented, i.e., $\Delta \mu $. Also,  those 40 points will be used to  reconstruct the function $\mu(z)$ via Gaussian process regression, as we describe next. We have used $\Omega_b/\Omega_m=0.157 \pm 0.002$ as given by Planck's results \cite{Planck2020}.

\section{Cosmological data}
\label{data}
\subsection{Gas Mass Fraction}
\begin{figure}[ht]
    \centering
    \includegraphics[scale=0.40]{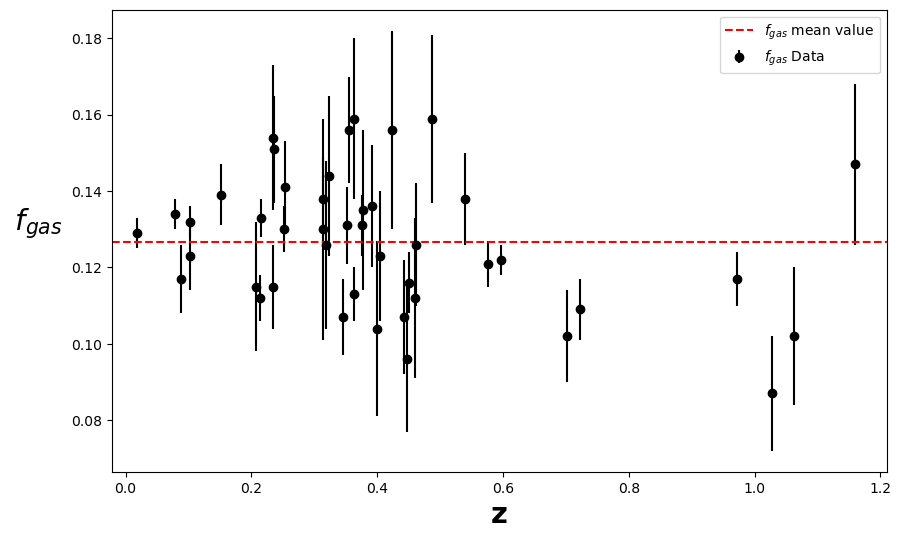}
    \caption{The 40 Chandra X-ray gas mass fraction as a function of redshift compiled by Mantz et al.~\cite{Mantz2014}.}
    \label{fgas_data}
\end{figure}

 We utilize the recent sample of 40 \(f_{\text{gas}}\) measurements spanning the redshift range \(0.018 \leq z \leq 1.160\), compiled in Ref.~\cite{Mantz2014}. This dataset comprises 40 massive, hot, and morphologically relaxed galaxy clusters observed by the Chandra telescope (see Fig.~\ref{fgas_data}). The selection of relaxed systems aims to minimize systematic biases in the hydrostatic masses. The gas mass fraction is computed within the spherical shell defined by \(0.25 \leq r/r_{2500} \leq 0.8\) rather than integrated over all radii. Further details about the data can be found in Refs.~\cite{Mantz2014}. 

 
%

\subsection{SNe Ia sample}
\begin{figure}[ht]
    \centering
    \includegraphics[scale=0.36]{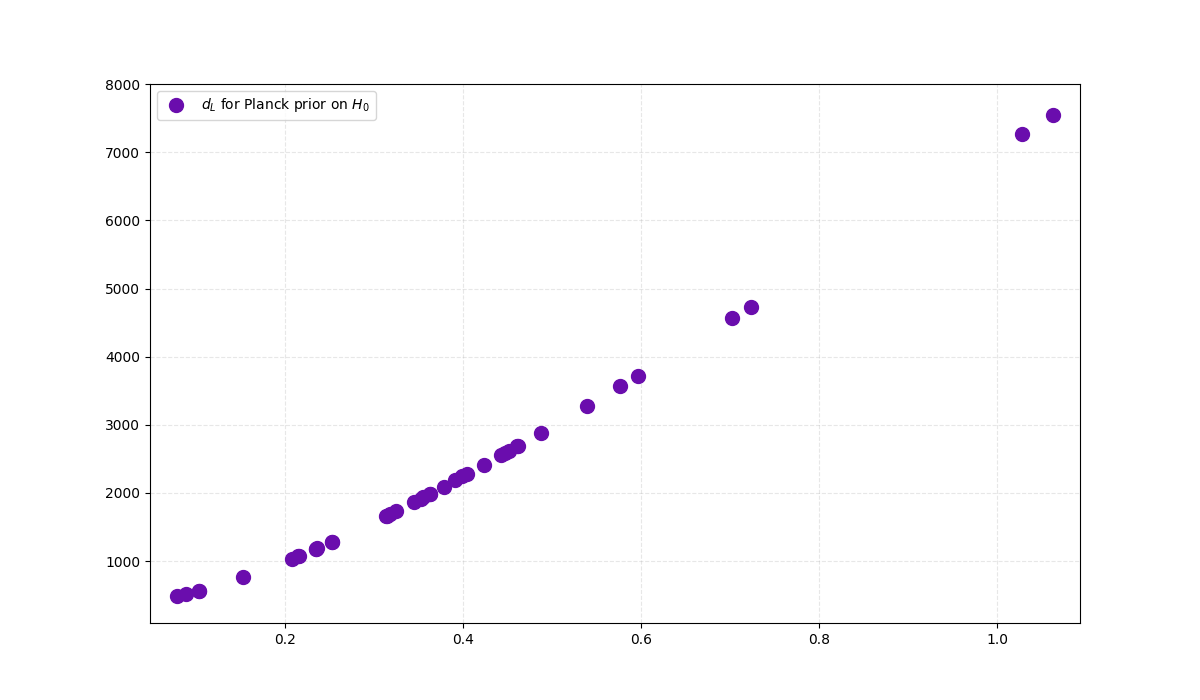}
    \caption{Luminosity distances for the 40 galaxy clusters obtained through gaussian process regression using the value for the Hubble constant, $H_{0}= 67.36 \pm 0.54$ km/s/Mpc, as given by Ref.~\cite{Planck2020}. }
    \label{gpr_dists}
\end{figure}

As formerly discussed, in order to perform our test, we need the luminosity distance for each galaxy cluster. To achieve this, we consider the PANTHEON dataset of Type Ia supernovae, compiled in 2018 by Scolnic et al.~\cite{Scolnic2018}, which combines observations from three subsets: 279 Type Ia supernovae from Pan-STARRS (\(0.03 < z < 0.68\)), samples from SDSS, SNLS, and various low-\(z\) samples from HST. This data set forms one of the largest combined sets of Type Ia supernovae, totaling 1048 events with redshift intervals ranging from \(0.01\) to \(2.26\). 
The luminosity distance (\(d_L\)) calculated for this sample is obtained by considering $H_{0}$ as provided by reference \cite{Planck2020} and as done in previous works~\cite{eu_rodrigo2} where $H_{0} = 67.36 \pm 0.54$ km/s/Mpc . To determine the luminosity distance for each galaxy cluster, we employ the Gaussian process methodology to compute the central value along with its associated variance. The results of the Gaussian process regression for \(d_L\) are shown in Fig.~\ref{gpr_dists}.

In the following sections, we present the main aspects of the 2HDM to be addressed in this work and translate the above analysis to constrain the \(\mu\) variation with redshift. This exemplifies how our approach can place bounds on Particle Physics models.

\section{The Two Higgs Doublet Model}
\label{2HDMsec}

 The 2HDM~\cite{2HDM_2012,Wang:2022yhm} consists of introducing an additional scalar to the SM spectrum, transforming as a doublet under the weak isospin group and carrying hypercharge $Y=+1$. As usual, we choose the vev of both doublets to be along the direction of weak isospin $T^3=-1/2$, the electrically neutral field components, so that expansion around the minimum energy configuration $\left<\Phi_i\right>_0$ reads, 
\begin{equation}
    \Phi_i=\left(\begin{array}{c}
         \phi_i^+  \\
         \phi_i^0
    \end{array}\right)=\left(\begin{array}{c}
         \phi_i^+  \\
         \frac{1}{\sqrt{2}}(v_i+\rho_i+i\eta_i) 
    \end{array}\right)\ ,\quad \left<\Phi_i\right>_0=\left(\begin{array}{c}
         0  \\
         v_i/\sqrt{2} 
    \end{array}\right)\ ,\quad i=1,2\,,
    \label{2HDM}
\end{equation}
where $\phi_i^+$ represents the electrically charged weak scalar eigenstates, while $\rho_i$ and $\eta_i$ are the neutral ones. The vevs are constrained by, $v=\sqrt{v_1^2+v_2^2}=246$~GeV, once both contribute to the gauge bosons' masses.
In order to obtain a simplified scalar potential, we introduce a discrete symmetry for the scalar doublets~\footnote{ This symmetry will soon be conveniently used to restrict the interaction of leptons with only one of the scalar multiplets.}, $Z_2:\ \Phi_1\rightarrow-\Phi_1\ ,\Phi_2\rightarrow\Phi_2$. Then, allowing only a bilinear,  $\Phi_1^\dagger\Phi_2+\text{h.c.}$ term, which softly breaks the $Z_2$ symmetry so as to avoid the domain wall problem~\cite{domain_wall}, the most general renormalizable, gauge, Lorentz and CP invariant scalar potential is written as, 
\begin{eqnarray}
V(\Phi_1,\Phi_2)&=&\displaystyle m_1^2\Phi_1^\dagger\Phi_1+m_2^2\Phi_2^\dagger\Phi_2-m_{12}^2(\Phi_1^\dagger\Phi_2+\Phi_2^\dagger\Phi_1)+\frac{\lambda_1}{2}(\Phi_1^\dagger\Phi_1)^2 
+\displaystyle\frac{\lambda_2}{2}(\Phi_2^\dagger\Phi_2)^2\nonumber  \\
&+&\lambda_3(\Phi_1^\dagger\Phi_1)(\Phi_2^\dagger\Phi_2)+\lambda_4(\Phi_1^\dagger\Phi_2)(\Phi_2^\dagger\Phi_1)
+\displaystyle\frac{\lambda_5}{2}\left[(\Phi_1^\dagger\Phi_2)^2+(\Phi_2^\dagger\Phi_1)^2\right]\,,
    \label{potential}
\end{eqnarray}
where $\lambda_j$ are dimensionless parameters and $m_1$, $m_2$ and $m_{12}$ possess mass dimension one. 
After the electroweak spontaneous symmetry breaking (SSB), we see that the scalars' mass terms in the potential, Eq.~(\ref{potential}), are not in a diagonal basis, so we need to rotate these fields to their mass eigenstates in order to obtain the mass spectrum of the model. From the original eight scalars, three of them are Goldstone bosons, $G$ and $G^\pm$, which are absorbed by the massive gauge bosons, $Z^0$ and $W^\pm$, respectively. The five remaining scalars get mixed to provide the mass eigenstates, namely, two CP-even scalars, $h$ and $H$, one CP-odd scalar, $A$ (these three scalars are electrically neutral), and two electrically charged ones, $H^\pm$. This physical basis (mass basis) is achieved through the following rotations~\cite{tessio},
\begin{equation}
\def\arraystretch{2.75}
\begin{array}{c}
\def\arraystretch{1.25}
    \left(\begin{array}{c}
         {h}  \\
         {H} 
    \end{array}\right)=R(\alpha)\left(\begin{array}{c}
         \rho_1  \\
         \rho_2 
    \end{array}\right)\ ,\quad \left(\begin{array}{c}
         {G}  \\
         {A} 
    \end{array}\right)=R(\beta)\left(\begin{array}{c}
         \eta_1  \\
         \eta_2 
    \end{array}\right)\ ,\quad \left(\begin{array}{c}
         G^\pm  \\
         H^\pm 
    \end{array}\right)=R(\beta)\left(\begin{array}{c}
         \phi_1^\pm  \\
         \phi_2^\pm 
    \end{array}\right)\,.
   \end{array}
\label{physical basis}
\end{equation}
Where, each of the rotation matrices, $R(\theta)$, with $\theta = \alpha \,,\beta$ can be written as,
\begin{eqnarray}
\begin{array}{c}
    R(\theta)\equiv\def\arraystretch{1}\left(\begin{array}{rr}
        \cos\theta & \sin\theta \\
        -\sin\theta & \cos\theta
    \end{array}\right)\, ,
    \end{array}
\end{eqnarray}
with 
\begin{equation}
tan(2\alpha)\equiv\displaystyle\frac{-2m_{12}^2v_1v_2+2\lambda_{345}v_1^2v_2^2}{m_{12}^2(v_2^2-v_1^2)+\lambda_1v_1^3v_2-\lambda_2v_1v_2^3}\,\,\,\,\,;\,\,\,\tan\beta\equiv\frac{v_2}{v_1}\,,   
\end{equation}
where we have defined $\lambda_{345}\equiv \lambda_3 + \lambda_4 + \lambda_5$.

The scalar's mass spectrum then becomes:
\begin{eqnarray}
    m^2_{h}&=&\displaystyle\frac12(\Delta_\rho-\delta_\rho)\,;\,\,\,\,\,\,\,\,
    m^2_{H}=\displaystyle\frac12(\Delta_\rho+\delta_\rho)\,; \,\,\,\,\,\,\,\,
    m^2_{A}=\displaystyle\left[\frac{m_{12}^2}{v_1v_2}-\lambda_5\right](v_1^2+v_2^2)\,; \\
    m^2_{H^\pm}&=&\displaystyle\left[\frac{m_{12}^2}{v_1v_2}-\frac{1}{2}(\lambda_4+\lambda_5)\right](v_1^2+v_2^2)\,,
\end{eqnarray}
where we have defined,
\begin{equation}
\Delta_\rho=\displaystyle\frac{m_{12}^2(v_1^2+v_2^2)}{v_1v_2}+\lambda_1v_1^2+\lambda_2v_2^2\,\,\,\,\,\mbox{and}
    \,\,\,\,\,\,\delta_\rho=\displaystyle\left[\left(\frac{m_{12}^2(v_1^2-v_2^2)}{v_1v_2}-\lambda_1v_1^2+\lambda_2v_2^2 \right)^2+4(m_{12}^2-\lambda_{345}v_1v_2)^2\right]^\frac{1}{2}\,.\nonumber
\end{equation}
In the above description, it becomes evident that the neutral CP-even mass hierarchy is $m_H > m_h$. 
For our purposes, the relevant interactions for future computations (electron mass, anomalous magnetic moment of the muon) come from the Yukawa Lagrangian,  
\begin{eqnarray}
-\cal{L}_\text{Y} &=& y_{ij}^{1e}(\bar{L}^i\Phi_1)e_R^j + y_{ij}^{1d}(\bar{Q}^i\Phi_1)d_R^j + y_{ij}^{1u}(\bar{Q}^i\tilde{\Phi}_1)u_R^j \nonumber \\
&+& y_{ij}^{2e}(\bar{L}^i\Phi_2)e_R^j + y_{ij}^{2d}(\bar{Q}^i\Phi_2)d_R^j + y_{ij}^{2u}(\bar{Q}^i\tilde{\Phi}_2)u_R^j + \text{h.c.}\,,
\label{yukawa}    
\end{eqnarray}
where each term is summed over the family indices $i$ and $j$ and we defined $\tilde{\Phi}_i\equiv i\sigma_2\Phi_i^*$. In this expression, $L_i$ and $Q_i$ represent the usual Standard Model (SM) left-handed lepton and quark doublets, respectively, while $e_R$, $u_R$, and $d_R$ denote their corresponding right-handed singlet partners. The parameters $y_{ij}$ denote the matrix elements (not necessarily diagonal) of the dimensionless Yukawa couplings.
 After the spontaneous symmetry breaking, this Lagrangian leads to the following fermionic mass matrix elements :
\begin{equation}
    M^f_{ij}=\displaystyle\frac{v_1}{\sqrt{2}}y_{ij}^{1f}+\frac{v_2}{\sqrt{2}}y_{ij}^{2f}\,,
    \label{nondiagonalmass}
\end{equation}
with the index $f$ denoting a quark or a lepton with the same weak isospin.

Observe that, differently from the SM, here we have two unrelated sources of contributions to fermion mass terms, one from the matrix $v_1 y^{1f}$ and another from $v_2 y^{2f}$. The bold implication is that there is no change of basis that diagonalizes both terms simultaneously, leading to Flavor Changing Neutral Currents (FCNC) in the fermion-scalar couplings. The fact that such interactions are experimentally unobserved calls for some arrangement to strongly suppress them. This can be achieved if we demand that each fermion mass matrix comes from only one of the scalar doublets, so that diagonalizing all fermion mass matrices, $M^f$, implies the automatic  diagonalization of $y^{1,2}$ (with the corresponding change from fermion interaction eigenstates to mass eigenstates) and, consequently, the disappearance of FCNC. This mechanism is known as Natural Flavor Conservation (NFC)~\cite{diag,diaglashow}. The easiest way to implement that is through the introduction of an extra symmetry, under which the scalar doublets transform differently, so that each fermion (also transforming under this symmetry) can couple to only one of the doublets. Our choice for that is a discrete $Z_2$ symmetry such that, $\Phi_1\rightarrow-\Phi_1\ ,\Phi_2\rightarrow\Phi_2$. This is the reason we omitted odd terms in either of the scalar doublets in the scalar potential, Eq.~(\ref{potential}), except for the soft $Z_2$ breaking term as we mentioned before.

The convention is to take all the fermion doublets, as well as singlet up-type quarks ($u_R$), to transform trivially, so that the up-type quarks always couple to $\Phi_2$ and we are left to decide which of the scalar doublets couples to the singlet down-type fermions (right-handed down-type quarks $d_R$ and charged leptons $e_R$).  In the scenario  we choose to work here, called  \textbf{lepton-specific} 2HDM (or LS2HDM for short), $d_R$ is assigned positive parity under $Z_2$ (so that both, $d_R$ and $u_R$ couple to $\Phi_2$, only) and $e_R$ is assigned negative parity (couples to $\Phi_1$, only). The main reason for choosing LS2HDM is because we restrict the electron to receive mass from one vev and the quarks from the other. In this way, we will have the simplest configuration regarding the variation of vevs with redshift (or, equivalently, $\tan{\beta}$), that is, we need to analyze only the case in which only one of the vevs varies, $v_1$, as a consequence of the variation of the gas mass fraction to be analyzed in the next section. We can then rewrite the Yukawa Lagrangian, Eq.~(\ref{yukawa}), in the mass basis for both, scalars and fermions~\cite{tessio},
\begin{eqnarray}
-\cal{L}_\text{Y} &=& \sum_{f=u,d,e} m_f\bar{f}f + \sum_{f=u,d,e}\frac{m_f}{v}(\xi_h^f \bar{f} f h + \xi_H^f \bar{f} f H -\imath \xi_A^f \bar{f}\gamma_5 f A)\nonumber \\
&-&\left[\frac{\sqrt{2}}{v}\bar{u}V_{ud}(m_u\xi_A^uP_L+m_d\xi_A^dP_R)dH^++\frac{\sqrt{2}m_e\xi_A^e}{v}\bar{\nu}_Le_RH^++\text{h.c.}\right],
\label{yukawa type X}
\end{eqnarray}
with,
\[
\xi_h^u = \xi_h^d=s_{\alpha}/s_{\beta}\ ,\ \xi_h^e=c_{\alpha}/c_{\beta}\ ,\ \xi_H^u=\xi_H^d=c_{\alpha}/s_{\beta}\ ,\ \xi_H^e=-s_{\alpha}/c_{\beta}\ ,\ \xi_A^u=-\xi_A^d=(t_{\beta})^{-1}\ ,\ \xi_A^e=t_{\beta}\,,
\]
where we defined $c_{\theta}\equiv\cos\theta$, $s_\theta\equiv\sin\theta$ and $t_{\theta}\equiv\tan\theta$ for shorthand notation, the sum is over all up-type and down-type fermions (leptons and quarks) and the coefficient $V_{ud}$ denotes the appropriate element of the CKM matrix. There is a linear combination of the neutral scalars that corresponds to the SM Higgs boson~\cite{tessio},
\begin{equation}
    H^\text{SM}\sim h\cos(\beta-\alpha)+H\sin(\beta-\alpha)\,
\end{equation}
which allows us to choose between $h$ and $H$ to represent the observed Higgs boson, $H^\text{SM}$, with mass around $125$~GeV, and consistent with SM predictions for its interactions with fermions, implying a constraint on the mixing angles $\alpha$ and $\beta$. Here we make the choice of $h$, the lighter scalar, to be the observed Higgs boson, so that $\alpha = \beta$, from now on. We are now prepared to obtain a constraint on this model from the bounds on $\mu$ varying with the redshift as inferred from the gas mass fraction in galaxy clusters, as we present next.

\section{Analysis and Results} 
\label{sec:analysis}

 Now we are going to obtain bounds on the variation of the electron to proton mass ratio by comparing the observed gas mass fraction within galaxy clusters to the theoretical gas mass fraction which is predicted by cosmological models as explained earlier in Sec. \ref{methodology}.
From the 40 data points we inferred a variation of $\mu$ ($\Delta \mu$), as follows: we take consecutive values of $\mu$ obtained from the data shown in table~\ref{tab:data_mu} (see the appendix) and subtract them $\Delta \mu = \mu_{j+1} - \mu_{j}$ so we have j-1 values of $\Delta \mu$, and the mean value is taken to be the expected variation of $\Delta \mu$ in this redshift span. We find that value to be:
\begin{equation}
\label{delta_mu_planck}
 \Delta \mu  = - 6.3 \cdot 10^{-3} \pm 2.7\cdot 10^{-1}.
\end{equation}

Although the constraints are not as restrictive as those obtained through quasar absorption spectra, the measurements are in agreement within 1 $\sigma$ confidence level. It is worth noting that the physics involved here is completely different. The techniques used to measure $\mu(z)$ in quasar spectroscopy rely on the hyperfine transition of molecules such as $\text{H}_2$, $\text{Fe}_{\text{II}}$, and $\text{Si}_{\text{II}}$ \cite{RevModPhys.88.021003, PhysRevD.72.043521}, with this transition being proportional to the ratio $\frac{\alpha_{EM}^{2} g_{p}}{\mu}$ and the effective radius of the nucleus. In this scenario, the ratio between the mass of the electron and the mass of the proton is not directly measured; rather, it is the ratio between the mass of the electron and the effective mass of the nucleus \cite{Martins_2017}. A measurement using the Eq.~(\ref{eqn_frac_gas_mod}) obtained through X-ray emission from ICM plasma, which consists predominantly of hydrogen ions, is a more direct measurement.
\begin{figure}[ht]
    \centering
    \includegraphics[scale=0.48]{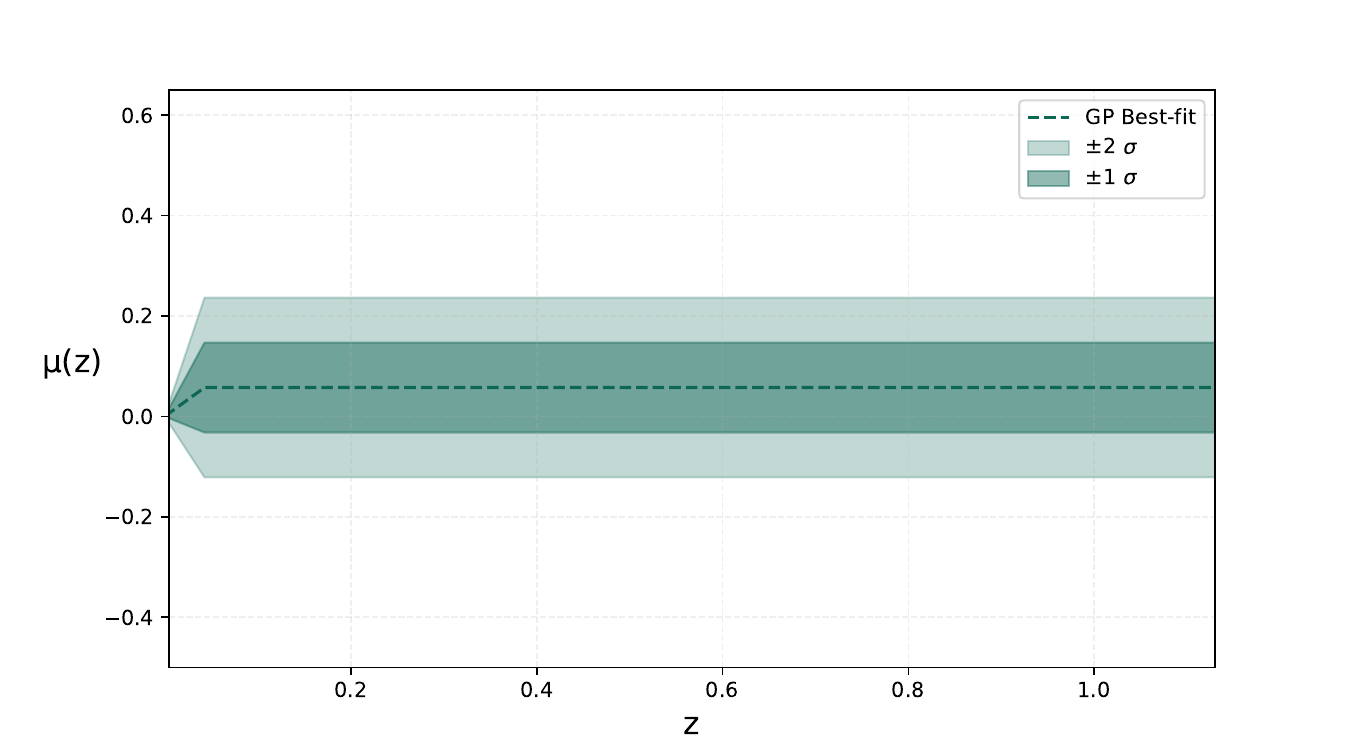}
    \caption{Reconstruction of $\mu(z)$ via gaussian processes regression by using the values from the Table II. The dashed line represents the best-fit curve, while the dark green region corresponds to the $1\sigma$ confidence interval, and the light green region represents the $2\sigma$ confidence interval.
}
    \label{gpr_besfit}
\end{figure}

Now, in order to obtain $\mu(z=0)$ from the observational values  shown in Table II, we apply a Gaussian process to those values (hereafter, we shall refer to it as $\mu_{\text{GPR}}(0)$). Gaussian Processes (GPs) offer a method for reconstructing a function from observational data without assuming a specific parametrization. The GP reconstruction process involves selecting a prior mean function and a covariance function that quantifies the correlation between the values of the dependent variable in the reconstruction, which are characterized by a set of hyperparameters such as the signal variance $\sigma$ and the characteristic length scale $l$. Unlike regular parameters, hyperparameters do not dictate the shape of a function; rather, they characterize typical variations in the reconstructed function. $l$ roughly represents the scale over which significant changes occur in the function values across $z$, while $\sigma$ indicates the typical magnitude of these alterations~\cite{seikel2013}\cite{scikit_learn_documentation}. While more complex covariance functions may involve additional hyperparameters, in this work we used a covariance function that solely incorporate $\sigma$ and $l$. To reconstruct $\mu(z)$, we do not specify a prior on $\mu$ to avoid any additional assumptions, we also optimize our kernel parameters via random search to avoid overfitting the model. We show the plot of our reconstruction in Fig.~\ref{gpr_besfit}  along with a Gaussian kernel serving as the covariance function, as follows,
\begin{equation}
    k(z,z') = \sigma^2 \exp{ \left( - \frac{(z - z')^2}{2l^2} \right)}.
\end{equation}
To perform the GP we used the Python scientific package scikit-learn~\cite{pedregosa2011}.  

\begin{figure}[ht]
    \centering
    \includegraphics[scale=0.51]{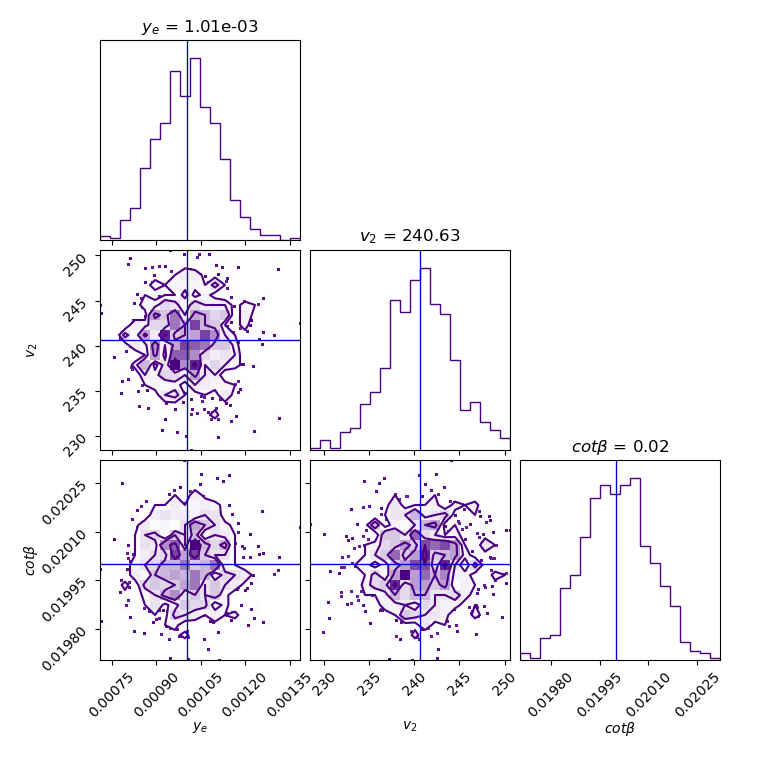}
    \caption{Constraints on $y_{e}$ , $v_{2}$ , and cot$\beta$ from $\mu(z)$ obtained using Plank's prior. The diagonal plots represent 1-D marginalized likelihood distributions of each parameter
present  and off-diagonal contours are 68\%, 95\%, and 99\% 2-D marginalized confidence regions.}
    \label{dist_values_PLANCK}
\end{figure}

Finally, we define a likelihood function that compares the values of \( \mu(z) \) obtained by GPR ($\mu_{GPR}$) taken when $z=0$, with the experimentally measured value (fiducial) \( \mu \) ($\mu_{\text{fid}}$),
\begin{equation}
\label{likelihood}
    L(y_e, v_2, cot\beta(z)) = \prod_{i=1}^{n} \frac{1}{\sqrt{2\pi\sigma_i^2}} \exp\left(-\frac{(\mu_{\text{GPR}}(z_i) - \mu_{\text{fid}}(z_i))^2}{2\sigma_i^2}\right)\,,
\end{equation}
to transition from the likelihood function \( L(y_e, v_2, \cot\beta(z)) \) to its logarithmic form 
\( \log L \), we focus on \( z = 0 \), where the likelihood reduces to a single comparison between the Gaussian Process Regression (GPR) prediction \( \mu_{\text{GPR}}(0) \) 
and the experimental value \( \mu_{\text{fid}}(0) \). The parameter \( \mu(z) \), defined as the ratio of the electron to proton mass in the two-Higgs-doublet model, incorporates \( \cot\beta(z) = v_1 / v_2 \), 
which is assumed to vary with \( z \). By fixing \( z = 0 \), 
the likelihood simplifies to a Gaussian form, allowing for a direct calculation of 
\( \log L \) for statistical purposes. 
\begin{equation}
\label{loglike}
\log L = -\frac{1}{2} \log(2\pi \sigma^2) - \frac{1}{2} \left( \frac{\mu_{\text{GPR}}(0) - \mu_{\text{fid}}(0)}{\sigma} \right)^2\,.
\end{equation}
We initialized the ensemble by choosing an initial value for the electron Yukawa coupling, \( y_{e} \), as 
\( 1 \times 10^{-3} \pm\) some gaussian noise of order $10^{-4}$. Similarly, the ratio of vevs in the LS2HDM, \( \cot\beta(z) \), was initialized as \( 1/50 \), with the noise term ensuring diversity in the initial values across the ensemble. Furthermore, \( v_{2} \) values were sampled from a Gaussian 
distribution centered at $240.5$. These initial conditions were then used to start the MCMC process.
Using the diagonalized Yukawa coupling for electrons in the LS2HDM, obtained from Eqs.~(\ref{yukawa},\ref{nondiagonalmass}), and assuming that any variation in $\mu$ with \(z\) arises from \(v_{1}(z)\), that is, the vev responsible for the electron mass, while the proton mass remains constant, we obtain,
\begin{equation}
\mu_{\text{fid}}(z) = \frac{y_{e}v_{2}}{\sqrt{2}m_{p}(\Lambda_\text{QCD})} \cot\beta(z). 
\label{eqnmuz}
\end{equation}
This expression, evaluated at \(z=0\), provides constraints for the Yukawa coupling of the electron, \(y_{e}\), for the cot$\beta_{0}$, and the vev, \(v_{2}\) we use in the log-likelihood from equation Eq.(\ref{loglike}).
The samples generated provide us with the best estimates for the parameters as given in Fig.~\ref{dist_values_PLANCK}, along with the conditional 2D regions for the confidence intervals. The results for the constrained parameters are summarized in  Table~\ref{tab:resumo_findings}.~\footnote{ We provide the source code for the replication of our analysis in the link for github repository: \href{https://github.com/aCosmicDebugger/2HDM}{2HDM}.}
\begin{table}[ht]
\centering
\renewcommand{\arraystretch}{1.5} 
\begin{tabular}{|c|c|}
\hline
\textbf{Parameter} & \textbf{Value} \\
\hline
$y_{e} \ $ & $(1.01\pm 0.07)\cdot 10^{-3}$ \\
\hline
$v_{2} \ (\text{GeV})$ & $240.57 \pm 2.93$ \\
\hline
$\cot\beta$ & $(2.003 \pm 0.081)\cdot 10^{-3}$ \\
\hline
\end{tabular}
\caption{Summary of the best values for the model's parameters obtained from our MCMC sampling.}
\label{tab:resumo_findings}
\end{table}

This constitutes the main result of our work. Next, we compute the contribution of LS2HDM to the muon anomalous magnetic moment and gather additional constraints to the model so as to find the favorable parameter space that is constrained by \(\mu(z)\).

\section{LS2HDM and the Muon anomalous magnetic moment}
\label{sec:muon-anomaly}

 There is, currently, a discrepancy between the muon anomalous magnetic moment,  $a_\mu\equiv (g-2)_\mu$, predicted by the SM and its experimental value, $\Delta a_\mu^{obs}\equiv a_\mu^\text{EXP}-a_\mu^\text{SM}=(25.1\pm5.9)\cdot10^{-10}$ \cite{muon(g-2),muon(g-2)/2}. We show here that our model is capable of accounting for that difference at $2\sigma$.

\begin{figure}[ht]
    \centering
    \includegraphics[scale=0.3]{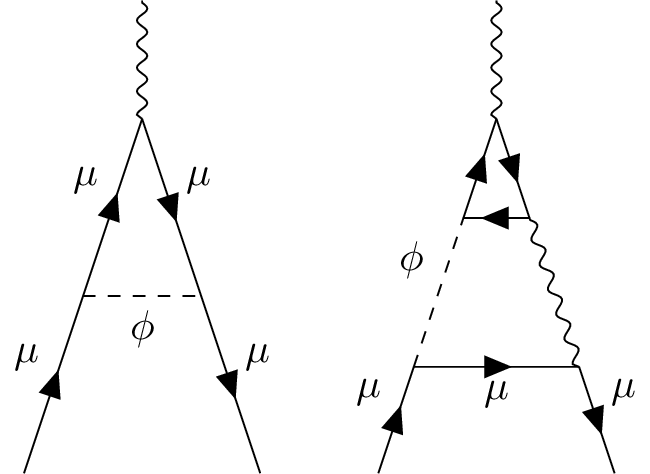}
    \caption{Neutral scalar contributions to muon, $(g-2)_\mu$ (the wiggy lines are photon ones).}
    \label{muon (g-2)}
\end{figure}

The most important contributions to $(g-2)_\mu$ from the LS2HDM are given by the diagrams in Fig.~\ref{muon (g-2)}, with $\phi$ representing the neutral scalars $h,\ H,\ A$. Here, $h$ is chosen to be the SM Higgs, with mass, $m_h=125$~GeV, whose contribution is already considered in the theoretical prediction so it does not enter in the computation of the deviation. Furthermore, the charged scalar contribution is insignificant, so it will not be considered here, meaning that we only take $H$ and $A$ into account. The first one loop diagram in Fig.~\ref{muon (g-2)} results in~\cite{Botella_2022},
\begin{equation}
    \Delta a_{\mu}^\text{1 loop}=\frac{1}{8\pi^2}\left(\frac{m_{\mu}}{v}\right)^2\left\{(\xi_H^\mu)^2\left[2I_1\left(\frac{m_{\mu}^2}{m_H^2}\right)-I_2\left(\frac{m_{\mu}^2}{m_H^2}\right)\right]-(\xi^\mu_A)^2I_2\left(\frac{m_{\mu}^2}{m_A^2}\right)\right\}\,,
\end{equation}
with,
\begin{eqnarray}
    I_1(x) &=& 1+\frac{1-2x}{2x\sqrt{1-4x}}\ln\left(\frac{1+\sqrt{1-4x}}{1-\sqrt{1-4x}}\right)+\frac{1}{2x}\ln{x} \nonumber \\
    &\approx& x\left(-\frac32 -\ln{x}\right)+x^2\left(-\frac{16}{3}-4\ln{x}\right)+{\cal O} (x^3) \,,
\end{eqnarray}
\begin{eqnarray}
    I_2(x) &=&\frac12+\frac{1}{x}+\frac{1-3x}{2x^2\sqrt{1-4x}}\ln\left(\frac{1+\sqrt{1-4x}}{1-\sqrt{1-4x}}\right)+\frac{1-2x}{2x^2}\ln x
    \nonumber \\
    &\approx& x\left(-\frac{11}{6} -\ln x\right)+x^2\left(-\frac{89}{12}-5\ln x\right)+{\cal O}(x^3) \,.
\end{eqnarray}

At one loop, $H$ contributes positively, and $A$ contributes 
 negatively~\cite{constraints_2022}, so that the overall contribution is not enough to explain the anomaly. For this reason we include the second diagram in Fig.~\ref{muon (g-2)}, a special type of two loop contribution called Barr-Zee diagram, which yields~\cite{Botella_2022},
\begin{equation}
    \Delta a_{\mu}^\text{Barr-Zee}=-\frac{2\alpha}{\pi}\frac{1}{8\pi^2}\left(\frac{m_{\mu}}{v}\right)^2\sum_f N_c^fQ_f^2\left\{\xi_H^\mu\xi_H^ff\left(\frac{m_f^2}{m_H^2}\right)-\xi_A^\mu\xi_A^fg\left(\frac{m_f^2}{m_A^2}\right)\right\}
\end{equation}
\begin{equation}
    f(z)=\frac{z}{2}\int_0^1dx\frac{1-2x(1-x)}{x(1-x)-z}\ln\left(\frac{x(1-x)}{z}\right)
\end{equation}
\begin{equation}
    g(z)=\frac{z}{2}\int_0^1dx\frac{1}{x(1-x)-z}\ln\left(\frac{x(1-x)}{z}\right)
\end{equation}

Here $H$ contributes negatively, and $A$ contributes positively~\cite{constraints_2022}. We sum over all the fermions, $f$, in the loop, each one characterized by its color number, 
$N_c^f$ (which is equal to 3 for quarks and zero for leptons), and its electric charge (multiple of unit positron charge), $Q_f$.

The above expressions depend on $m_H$, $m_A$ and $\tan{\beta}$, which are in principle hard to constrain because (considering $\beta=\alpha$) the quark couplings from Eq.~(\ref{yukawa type X}) are proportional to $\cot\beta\ll 1$. This means that hadronic cross sections are attenuated and thus incapable of providing bounds for the 2HDM masses. So it becomes necessary to establish model-dependent bounds on these parameters, and it is this feature that reveals the importance of new fronts to obtain these bounds, like the electron varying mass inferred from the electron to proton mass ratio, $\mu$, discussed in the previous sections. We follow here the model suggested by Ref.~\cite{constraints_2022}, which uses lepton universality constraints, the CDF W boson mass results and muonic $g-2$ measurements to probe the allowed parameter space (as well as the requirement of vacuum stability, perturbativity and unitarity in the wrong sign scenario). The $h\rightarrow AA$ decay channel is considered, adding the constraint $m_h\ge 2m_A$.

\begin{figure}[ht]
    \centering
    \includegraphics[scale=.12]{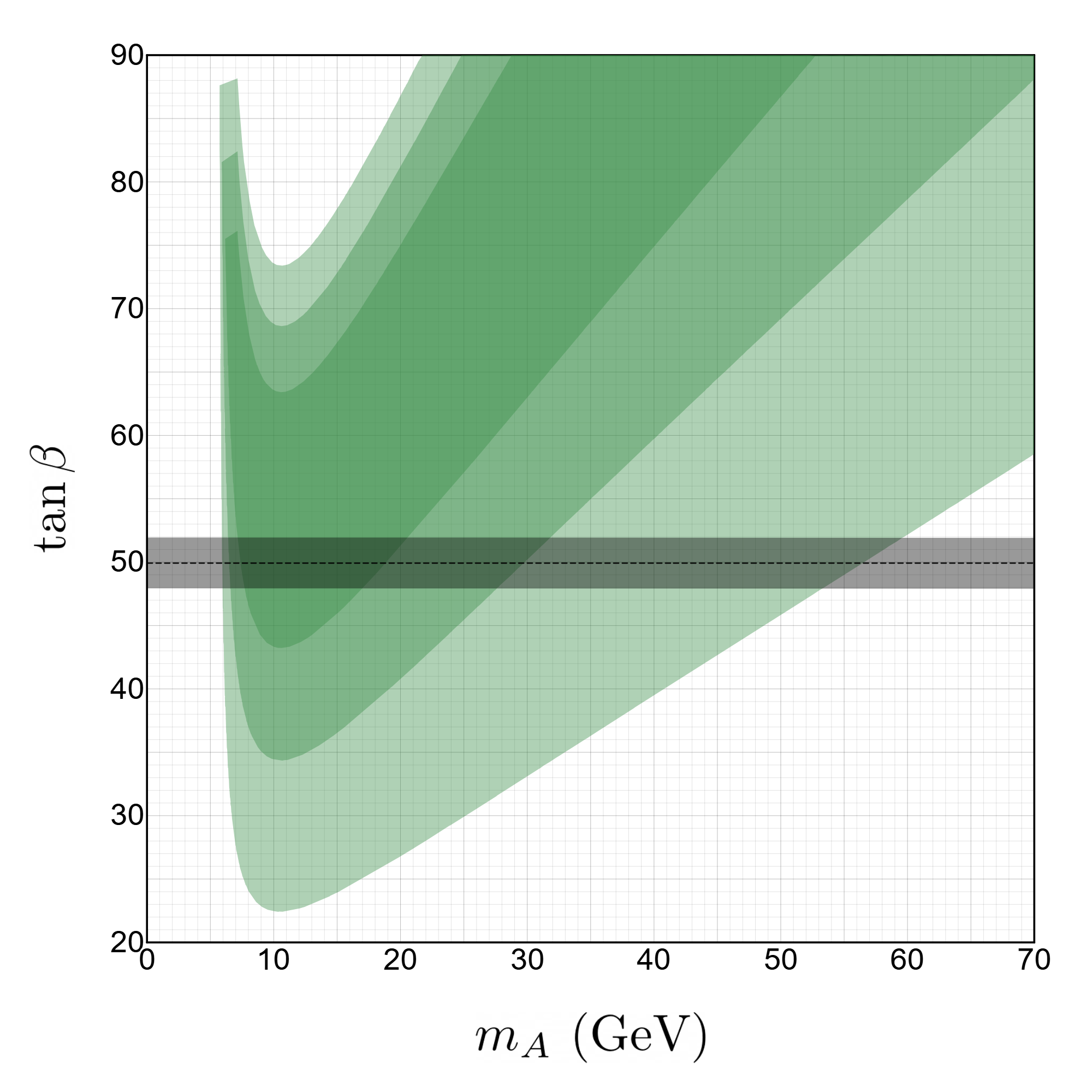}
    \caption{Constraints on 2HDM parameters $m_A$ and $\tan{\beta}$ using the $(g-2)_\mu$ prediction of LS2HDM ($m_H$ is fixed at 300 GeV). The green regions represent, from darker to lighter, $1\sigma$, $2\sigma$ and $3\sigma$, where a $\chi^2$ distribution is adopted. The region in light gray represents $\cot\beta=0.02003 \pm 0.00081$ from Table~\ref{tab:resumo_findings}.}      
    \label{tbeta mA}
\end{figure}

\begin{figure}[ht]
    \centering
    \includegraphics[scale=.1]{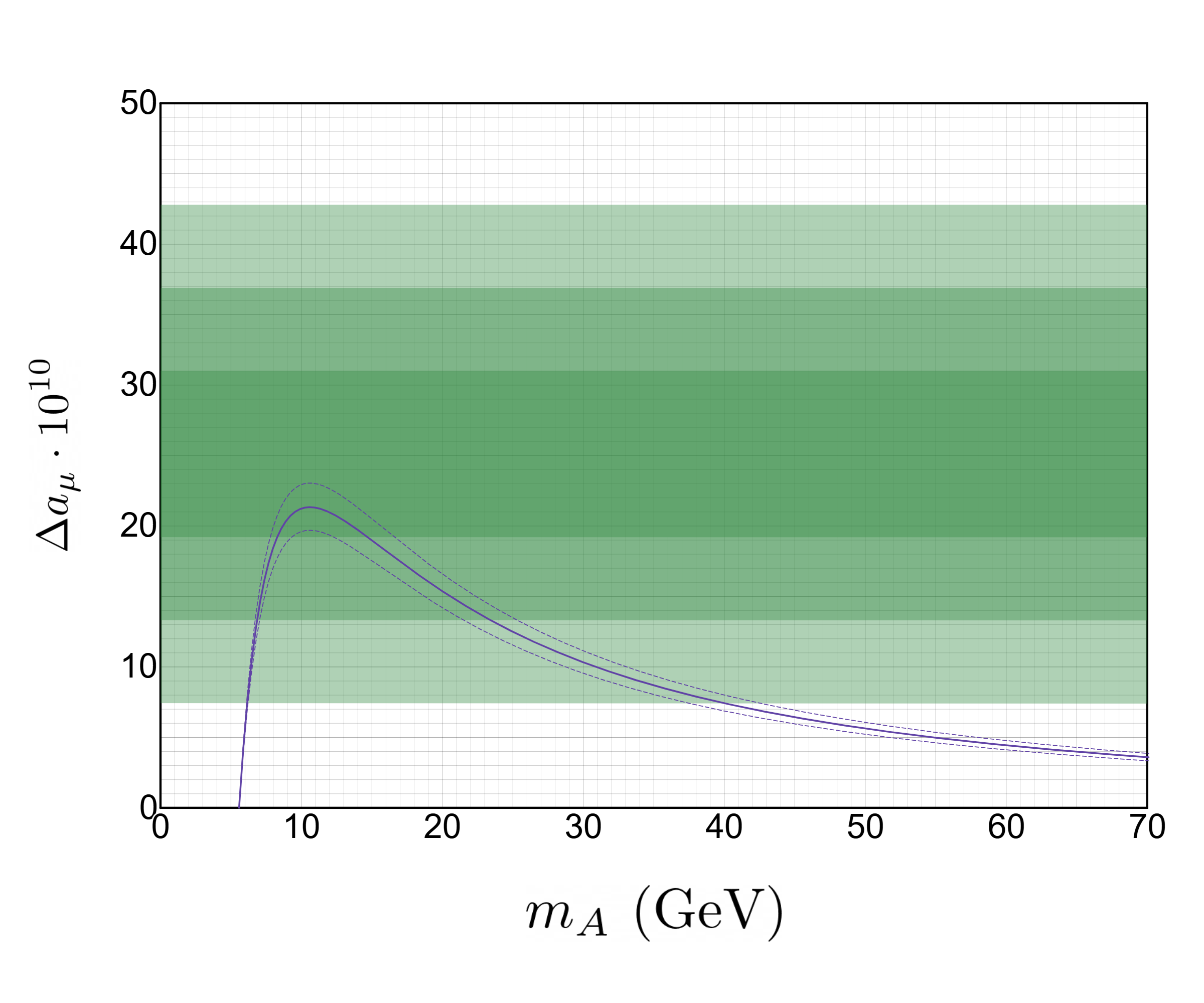}
    \caption{$\Delta a_\mu^\text{2HDM}$ as a function of $m_A$ with $m_H=300$ GeV. We took the central value $\tan{\beta}=49.92$ (solid purple curve) and the limiting values at $1\sigma_{\tan\beta}=2.02$ (dashed purple curves). The green bands are the observed $(g-2)_\mu$ discrepancy at $1\sigma$, $2\sigma$ and $3\sigma$.}
    \label{mH mA}
\end{figure}

The regions in parameter space that meet these requirements and keep $\Delta a_\mu^\text{2HDM}(m_H,m_A,t_{\beta})=\Delta a_\mu^\text{1 loop}+\Delta a_\mu^\text{Barr-Zee}$ inside $1\sigma$, $2\sigma$ or $3\sigma$ deviation from $\Delta a_\mu^{obs}=(25.1\pm 5.9)\cdot 10^{-10}$ are shown in Figs.~\ref{tbeta mA} and (\ref{mH mA}), displayed as green bands. In Fig.~\ref{tbeta mA}, we present the predictions of $\tan{\beta}$ against the pseudoscalar mass, $m_A$, keeping the mass of the heavier scalar, $m_H$, fixed at $m_H = 300$~GeV. We take into account the bounds stemming from our astrophysical analysis of $\mu(z)$ in Table~\ref{tab:resumo_findings}, representing $\cot\beta=0.02003 \pm 0.00081$  (light gray band). We see that a light pseudoscalar is favored, with mas ranging from around 10 to 50~GeV. As for the Fig.~\ref{mH mA}, our prediction for $\Delta a_\mu^\text{2HDM}$ is plotted as a function of $m_A$ with $m_H=300$ GeV. We kept  $\tan{\beta}$ fixed with central value $\tan{\beta}=49.92$ (solid purple curve) and the limiting values at $\pm 1\sigma_{\tan\beta}=\pm 2.02$ (dashed purple curves). The green bands are the observed $(g-2)_\mu$ discrepancy at $1\sigma$, $2\sigma$ and $3\sigma$. From this figure we conclude that the neutral scalar contributions can solve the muon anomaly at $1\sigma$ when $m_A\gtrapprox 10$~GeV. The critical point at $m_A\approx 10.67$~GeV is due to the competition between the one loop (negative) and Barr-Zee (positive) contributions.

\begin{figure}[ht]
    \centering
    \begin{tikzpicture}
        \node (c) {\includegraphics[scale=.09]{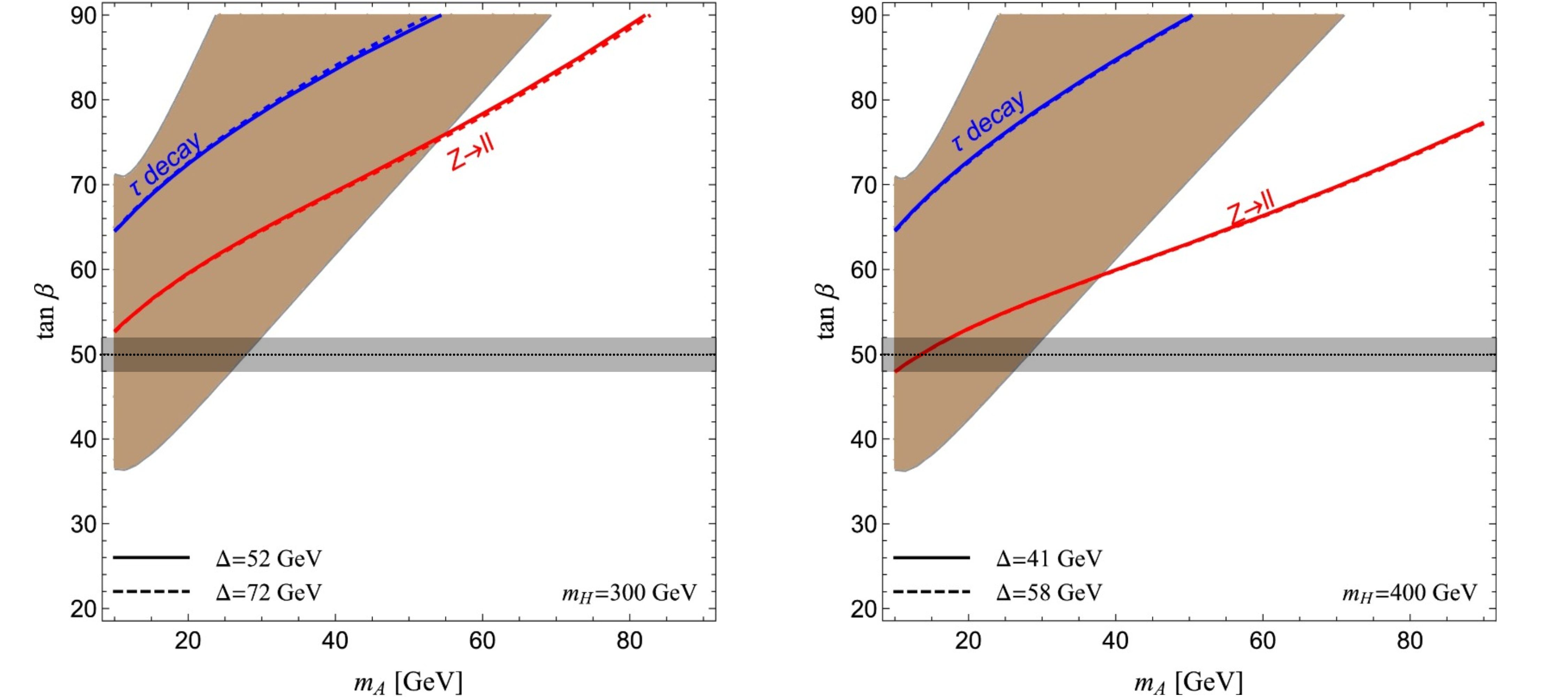}};
        \node[above=0.35cm of c] (u) {};
        \node[left=3cm of u] (a) {(a)};
        \node[right=3.25cm of u] (b) {(b)};
        \node[below=0cm of c] (d) {};
        \draw[-] (u) -- (d);
    \end{tikzpicture}
    \caption{(a) - (b) Constraints given by \cite{constraints_2022}. The $2\sigma$ region allowed by the muon $g-2$ is sketched in brown. The colored lines give upper bounds on $\tan\beta$ from lepton universality tests, exploring both $\tau$ decay and $Z$ decay in any pair of leptons. The full and dashed lines  are associated with the lower and upper bounds on $\Delta\equiv m_{H^\pm}-m_H$, respectively, and we superimposed the region in light gray to represent our astrophysical results. In Fig.~\ref{kim}a, $m_H$ is fixed at 300 GeV and $\Delta$ ranges from 52 to 72 GeV. In Fig.~\ref{kim}b, $m_H$ is fixed at 400 GeV and $\Delta$ ranges from 41 to 58 GeV.}
    \label{kim}
\end{figure}
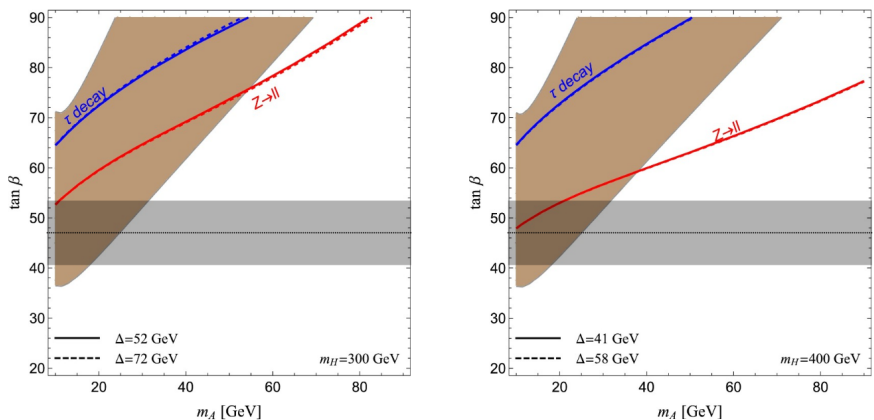

Just for completeness, we use some additional constraints from Ref.~\cite{constraints_2022} that give $m_{H^\pm}\gtrapprox m_H$, besides solving the $(g-2)_\mu$ discrepancy and the W mass anomaly simultaneously. For that, we superpose our results from the $\mu(z)$ bounds on $\tan{\beta}$ in the LS2HDM (table~\ref{tab:resumo_findings}) to their results on   $(g-2)_\mu$, universality tests through $\tau$-decay and $Z\rightarrow \bar{l} l$ (Z decay into a pair of leptons), as well as the scalar mass difference, $\Delta\equiv m_{H^\pm}-m_H$, constrained by the requirement that it solves the $W$ mass anomaly~\cite{CDF:2022hxs}.
We see from Fig.~\ref{kim} that these extra bounds are compatible with our constraints on $\tan{\beta}$ in the $1\sigma$ range $47.9 \leq \tan{\beta}\leq 51.9$, covering the central value obtained before with a small variation depending on the heavier neutral scalar mass. The first panel in that figure is for a heavier Higgs, $m_H = 300$~GeV, while the second one is for $m_H=400$~GeV. The upper bounds on $\tan{\beta}$ from $\tau$-decay and $Z\rightarrow \bar{l}l$ are shown as full and dashed lines for different choices of the scalar mass difference, $\Delta$. Observe that their results for the $(g-2)_\mu$, sketched in brown, is a little bit different from ours because they use only the lepton $\tau$ in the Barr-Zee loop, while we summed over all the SM fermions.
%

\section{Conclusions}
\label{sec:conclusions}
 In this work, we analyze the expression for the gas mass fraction of a galaxy cluster and found that \(f_{\text{gas}} \propto \left(1+\mu\right)^{2}\). Thus, by combining Type Ia Supernovae observations with gas mass fraction measurement data from 40 galaxy clusters, we imposed constraints on a possible variation of the electron-to-proton mass ratio, \(\Delta \mu\), as given in Eq.~(\ref{delta_mu_planck}). Additionally, using a Gaussian regression process, we reconstructed a \(\mu(z)\) curve, as given in Eq.~(\ref{eqnmuz}), which allows us to impose limits on models beyond the electroweak Standard Model (SM) of particle physics. 

To illustrate our approach, we employed a specific extension of the scalar sector of the SM by adding an extra Higgs doublet and an additional discrete \(Z_2\) symmetry. This resulted in the two-Higgs doublet model in the scenario of lepton-specific coupling, which we referred to as LS2HDM, presented in Sec.~\ref{2HDMsec}. This choice was made not only for simplicity but also to give meaningful context to a well-studied class of models in the literature. We were able to translate the results on \(\mu(z)\) into constraints on the parameters of the model, namely \(\cot{\beta}\) (reverted to \(\tan{\beta}\)), \(v_2\), and \(y_e\) (see Table~\ref{tab:resumo_findings}), finding their optimal values through MCMC sampling.

These constraints, when further compared with the model's ability to account for the discrepancy between experimental and SM theoretical values of the anomalous muon magnetic moment, \((g-2)_\mu\), resulted in a robust set of bounds on the parameters. This analysis favored a relatively light pseudoscalar, with \(m_A \gtrapprox 10~\text{GeV}\), and a heavier scalar with mass around \(300 \leq m_H \leq 400~\text{GeV}\). These conclusions were obtained by combining existing bounds~\cite{constraints_2022} on the model, including constraints from \(\tau\)-decay and \(Z \rightarrow \bar{\ell}\ell\), along with the accommodation of the W-mass anomaly observed by the CDF collaboration~\cite{CDF:2022hxs}. 

Finally, we emphasize that the main goal of this work was to explore a complementary approach to Particle Physics models at the interface with Astrophysics and Cosmology. In our case, this approach was derived from the properties of gas in galaxy clusters. Our results highlight the importance of seeking new constraints for non-standard physics in environments beyond particle physics accelerators, as can be readily seen in Figs.~\ref{kim}a and \ref{kim}b, where the allowed parameter space has been further reduced to the gray range when our astrophysical analysis is taken into account. Admittedly, we restricted ourselves to a very specific model, but it served our purpose of demonstrating that new paths can be followed to unite these areas, without evident correlation, with great gain for model building in Particle Physics, confirming that the large-scale structures of the universe may contain more information than previously anticipated. In the coming years, there will be a significant increase in the quantity and quality of observations made possible by instruments such as the eROSITA telescope, a Russian-German collaboration. This progress will allow more comprehensive statistical analyses, making it feasible to reach more restrictive limits using the methodology described in this paper.

\acknowledgments
PSRS would like to thank the financial support from  Conselho
Nacional de Desenvolvimento Cientifico e Tecnologico
(CNPq) under the project No. 309650/2019-4 and the support of UFPB through an internal project PIA16656-2023; RFLH thanks the financial support from the Conselho
Nacional de Desenvolvimento Cientıfico e Tecnologico
(CNPq) under the project No. 308550/2023-47; RGA thanks the financial support from UFPB through PIBIC-CNPq under the project No. PIA16656-2023.

\newpage

\appendix
\section{40 observations of $\mu$}

The function \texttt{delta\_mu\_with\_error} in \href{https://github.com/aCosmicDebugger/2HDM}{2HDM} calculates the differences between the values of $\mu$ at consecutive positions and their respective uncertainties. The formula for each $\Delta \mu$ is given by: $\Delta \mu_i = \mu_{i+1} - \mu_i$. And for the associated error of each $\Delta \mu_i$, we have  $\Delta \sigma_{\mu_i} = \sqrt{\sigma_{\mu_{i+1}}^2 + \sigma_{\mu_i}^2}$. Finally, for the $\Delta \mu \equiv \langle \Delta \mu \rangle$
\begin{equation}
    \langle \Delta \mu \rangle = \frac{1}{n} \sum_{i=1}^{n} \Delta \mu_i
\end{equation}
where \( \Delta \mu_i \) are the differences of $\mu$ and \( n \) is the total number of values. We use the $\langle \Delta \mu \rangle $ as our estimate for $\Delta \mu$ in Eq. (\ref{delta_mu_planck})

\begin{table}[h!]
\centering
\begin{tabular}{|>{\raggedright\arraybackslash}p{2.5cm}|>{\raggedright\arraybackslash}p{2.5cm}|}
\hline
\textbf{$\mu$} & \textbf{$\sigma_{\mu}$} \\ \hline
0.078   & 0.259413  \\ \hline
0.088   & 0.203252  \\ \hline
0.103   & 0.115985  \\ \hline
0.103   & 0.143774  \\ \hline
0.152   & 0.095311  \\ \hline
0.208   & -0.015405 \\ \hline
0.214   & -0.025265 \\ \hline
0.215   & -0.029885 \\ \hline
0.235   & 0.102452  \\ \hline
0.235   & 0.050020  \\ \hline
0.236   & 0.062295  \\ \hline
0.252   & 0.020226  \\ \hline
0.253   & 0.109813  \\ \hline
0.313   & 0.093470  \\ \hline
0.314   & 0.044570  \\ \hline
0.318   & 0.069704  \\ \hline
0.324   & 0.078395  \\ \hline
0.345   & -0.044600 \\ \hline
0.352   & -0.007903 \\ \hline
0.355   & 0.129140  \\ \hline
0.363   & 0.156912  \\ \hline
0.363   & -0.024807 \\ \hline
0.378   & 0.103819  \\ \hline
0.391   & 0.080763  \\ \hline
0.399   & -0.087368 \\ \hline
0.404   & -0.021369 \\ \hline
0.423   & 0.215044  \\ \hline
0.442   & -0.074786 \\ \hline
0.447   & -0.079323 \\ \hline
0.451   & 0.013216  \\ \hline
0.460   & 0.013755  \\ \hline
0.461   & 0.039923  \\ \hline
0.487   & 0.135969  \\ \hline
0.539   & 0.134061  \\ \hline
0.576   & 0.043878  \\ \hline
0.596   & 0.189276  \\ \hline
0.702   & -0.146335 \\ \hline
0.723   & 0.184268  \\ \hline
1.028   & 0.013755  \\ \hline
1.063   & 0.011109  \\ \hline
\end{tabular}
\caption{Data from 40 observations of $\mu$ as obtained from Eq. (\ref{fgas_mu})}
\label{tab:data_mu}
\end{table}

\newpage

\bibliography{ref}

\end{document}